\documentclass[12pt]{iopart}
\usepackage{iopams, setstack, amssymb} 

\begin{document}
\title{Geometric quenches in quantum integrable systems}

\author{Jorn Mossel, Guillaume Palacios and Jean-S\'ebastien Caux }
\address{Institute for Theoretical Physics, University of Amsterdam, Science Park 904, Postbus 94485, 1090 GL Amsterdam, The Netherlands}

\begin{abstract}
We consider the generic problem of suddenly changing the geometry of an integrable, one-dimensional many-body quantum system.  
We show how the physics of an initial quantum state released into a bigger system can be completely described within the framework 
of the Algebraic Bethe Ansatz, by providing an exact decomposition of the initial state into the eigenstate basis of the system after
such a geometric quench.  Our results, applicable to a large class of models including the Lieb-Liniger gas and Heisenberg spin chains, 
thus offer a reliable framework for the calculation of time-dependent expectation 
values and correlations in this nonequilibrium situation.
\end{abstract}

\maketitle

%\section{Introduction}

Integrable models have provided a wealth of valuable information on the equilibrium thermodynamics of strongly correlated systems 
during the last decades (see \cite{GaudinBOOK,Korepin1993,TakahashiBOOK} and references therein).
More recently, much progress has also been made on their dynamical properties.  
On the other hand, the field of out-of-equilibrium dynamics is only now starting to become accessible to comparably exact methods \cite{Faribault2009, Rossini2009, Gritsev2010, Fioretto2010, Mossel2010}, in particular in the study of so-called quantum quenches (see \cite{Calabrese2006}-\cite{NJP2010} for example). In those studies the quench is defined as a sudden change of a local parameter of the Hamiltonian, typically the interaction strength or an external field. In this short note, we consider the case where a global parameter is abruptly changed, namely the size of the system before and after the quench. We baptize this the geometric quench. Earlier  studies of such a setup, using numerical techniques, include \cite{Rigol2007,Heidrich-Meisner2008}.

 In this paper, we highlight an interesting feature of the wavefunctions of integrable one-dimensional models of quantum many-body physics,
opening the door to a large class of possible calculations of nonequilibrium dynamical effects in these systems upon a sudden change of their geometry. We start by a simple explanation using the coordinate representation of wavefunctions, and provide an algebraic version of the
reasoning thereafter.  The paper ends with some perspectives and suggestions for further work.

\section{Coordinate Bethe Ansatz}
We consider a quantum system initially defined within a finite space interval $0 \leq x < L_1$
(for clarity, we use notations appropriate for a model defined on the continuum, such as the Lieb-Liniger model \cite{1963_Lieb_PR_130_1};  our reasoning also
applies to lattice models such as spin chains by trivial modifications).  
We suppose that the initial Hamiltonian can be written as the integral of a Hamiltonian density
\begin{equation}
H_1 = \int_0^{L_1} \mathrm{d}x\, {\cal H} (x)
\end{equation}
and that specific boundary conditions ({\it e.g.} periodic or twisted;  the reasoning for the open case is slightly different, and is discussed in the next section) labeled by $\alpha_1$ are imposed on the wavefunctions.
In the case of integrable systems, a complete set of orthonormalizable eigenfunctions is explicitly given 
in each $N$-particle sector by a Bethe Ansatz;  individual wavefunctions
\begin{equation}
\Psi^{(1)} (\{ x \} | \{ \lambda \}^{L_1}_{\alpha_1}), \hspace{1cm} 0 \leq x_i < L_1
\end{equation}
are labeled by their cardinality-$N$ set $\{ \lambda \}^{L_1}_{ \alpha_1}$ of rapidities, solution to the Bethe equations associated to 
quantization in the space $0 \leq x < L_1$ under boundary conditions $\alpha_1$.
Completeness ensures that the number of allowable solutions coincides with the dimensionality of the Hilbert space.  One important observation,
reflected in our notation, is that the eigenfunctions $\Psi (\{ x \} | \{ \lambda \} )$ do not explicitly depend 
on the length $L_1$ or boundary conditions $\alpha_1$;  rather, these dependencies are carried only implicitly by the
sets of rapidities.  This simple observation forms the basis of what we propose here.

The generic experiment we have in mind can be described as follows.
Suppose that we initially prepare the system in a specific eigenstate 
$\Psi (\{ x \}, t\!\! =\!\! 0^-) \equiv \Psi^{(1)} (\{ x \} | \{ \lambda \}^{L_1}_{\alpha_1})$.
At time $t = 0$, we glue a new, empty region onto the original system, thereby changing its size from $L_1$ to $L_2 > L_1$, a process we 
call a geometric quench.  From this point onwards, to control the subsequent time evolution exactly, 
all calculations must be done in the basis of eigenstates of the new Hamiltonian
\begin{equation}
H_2 = \int_0^{L_2} \mathrm{d}x\, {\cal H} (x)
\end{equation}
defined by integrating the same Hamiltonian density as before (so all interaction parameters remain unchanged) over 
the larger space interval up to $L_2$.  The eigenstates of $H_2$ 
(choosing boundary conditions labeled by $\alpha_2$) are functions of the form
\begin{equation}
\Psi^{(2)} (\{ x \} | \{ \mu \}^{L_2}_{\alpha_2}), \hspace{1cm} 0 \leq x_i < L_2,
\end{equation}
where the sets of rapidities $\{ \mu \}^{L_2}_{\alpha_2}$ now solve the Bethe equations associated to quantization
in the space $0 \leq x < L_2$ under boundary conditions $\alpha_2$.  

At first sight, the gluing of the wavefunctions before and after the geometric quench is not simple.  This gluing condition
can be written as the nonlinear condition
\begin{eqnarray}
\Psi (\{ x \}, t = 0^+) 
%\Psi_c^{(1)} (\{ x \}| \{ \lambda \}_{(L_1, \alpha_1)}) 
= \left\{ \begin{array}{cc}
\Psi^{(1)} (\{ x \}| \{ \lambda \}^{L_1}_{ \alpha_1}), & 0 \leq x_i < L_1, \\
0 & \mbox{if}~ \exists ~x_i \in [L_1, L_2[. \end{array} \right. 
\label{eq:gluing_condition}
\end{eqnarray}
Decomposing this into the basis of eigenstates of $H_2$ can now be achieved by relying on 
the crucial observation made above, namely 
that the wavefunctions of $H_1$ (respectively $H_2$) do not depend explicitly on $L_1, \alpha_1$
(respectively $L_2, \alpha_2$).  This allows us to make the identifications
\begin{eqnarray}
\Psi^{(2)} (\{ x \} | \{ \mu \}^{L_2}_{\alpha_2})|_{0 \leq x_i < L_1} = \Psi^{(1)} (\{ x \} | \{ \mu \}^{L_2}_{\alpha_2} )|_{0 \leq x_i < L_1}.
\end{eqnarray}
In other words, in the region of space where the initial wavefunction is nonvanishing, we can exactly match the functional
form of Bethe wavefunctions of $H_2$ when restricted to the evaluation domain $0 \leq x_i < L_1$, to the wavefunctions of $H_1$ but this time
evaluated using the rapidities of the $\Psi^{(2)}$ state.  The overlap of the initial state with a given eigenfunction of $H_2$ 
is thus given by the well-known Slavnov formula \cite{Slavnov1989} (associated to the original domain),
\begin{eqnarray}
\hspace{-2.0cm} \frac{\langle \Psi^{(1)} (\{ \mu \}^{L_2}_{\alpha_2}) | \Psi^{(1)} ( \{ \lambda \}^{L_1}_{\alpha_1}) \rangle}
{\sqrt{ \langle \Psi^{(1)} ( \{ \lambda \}^{L_1}_{ \alpha_1}) | \Psi^{(1)} ( \{ \lambda \}^{L_1}_{ \alpha_1}) \rangle
\langle \Psi^{(2)} (\{ \mu \}^{L_2}_{ \alpha_2}) | \Psi^{(2)} (\{ \mu \}^{L_2}_{\alpha_2}) \rangle}} = F_1 ( \{ \mu \}^{L_2}_{\alpha_2}; \{ \lambda \}^{L_1}_{\alpha_1} )\, .
\label{overlapF1}
\end{eqnarray}
The exact decomposition of our initial state onto the wavefunctions of $H_2$ is simply written as
\begin{eqnarray}
\Psi (\{ x \}, t = 0^+) = \sum_{\{ \mu \}^{L_2}_{\alpha_2} } F_1 ( \{ \mu \}^{L_2}_{\alpha_2}; \{ \lambda \}^{L_1}_{\alpha_1} ) \Psi^{(2)} (\{ x \} | \{ \mu \}^{L_2}_{\alpha_2}).
\end{eqnarray}
The subsequent time evolution is now trivial to build back in, since $\Psi^{(2)}$ are exact eigenstates of $H_2$.
This simple expression is the central result of our paper.  

\section{Algebraic Bethe Ansatz}
\subsection{Geometric quench overlap: Periodic boundary conditions}
 From the  previous analysis  we clearly see that the 
the dynamics of the geometric quench problem is encoded in the overlaps $F_1$ defined by 
\eref{overlapF1}. The natural framework for computing norms and overlaps of Bethe eigenstates is the Algebraic Bethe Ansatz. This approach is based on the construction of a monodromy matrix $T(\lambda)$, which leads to an algebraic formulation of the problem, see for instance  \cite{Korepin1993}. The monodromy matrix, $T \in V_0 \otimes \mathcal{H}$ with $V_0$ an auxiliary space and $\mathcal{H}$ the Hilbert space, should satisfy the intertwining relation
\begin{equation}\label{intertwining}
R(\lambda-\mu) \left( T(\lambda) \otimes T(\mu) \right) = \left( T(\mu) \otimes T(\lambda) \right) R(\lambda-\mu) 
\end{equation}
where $R(\lambda)\in V_0 \otimes V_0$ is the $R-$matrix satisfying the Yang-Baxter equation. One can define the transfer matrix as $\mathcal{T}(\lambda) = \mbox{Tr}_0 T(\lambda)$, where the trace is taken over the auxiliary space. The transfer matrix  now acts as a generating function for  a set of mutually commuting conserved charges, including the Hamiltonian. For simplicity we focus in this discussion on XXZ-type models (which includes the Lieb-Liniger gas) given by an  $R$-matrix of the form
\begin{equation}
R(\lambda)=\left( \begin{array}{cccc}
f(\lambda) &0&0&0\\
0&g(\lambda)&h(\lambda)&0\\
0&h(\lambda)&g(\lambda)&0\\
0&0&0& f(\lambda)\\
\end{array}\right).
 \end{equation}
For this type of $R$-matrix the monodromy matrix $T(\lambda)$ is a $2\times 2$ matrix in auxiliary space,
\begin{equation}
T(\lambda) = \left( \begin{array}{cc} A(\lambda) & B(\lambda)\\ C(\lambda)&D(\lambda)
\end{array} \right).
\end{equation}
The commutation relations of the operators $A(\lambda),B(\lambda),C(\lambda),D(\lambda)\in \mathcal{H}$  follow from \eref{intertwining}. Using these operators, eigenstates of the transfer matrix (and thus of the Hamiltonian) can be constructed as
\begin{equation}
\langle \Psi| = \langle 0 | \prod_{j=1}^N C(\lambda_j) \qquad |\Psi\rangle = \prod_{j=1}^N B(\lambda_j) |0\rangle
\end{equation}
were the rapidities $\{\lambda_j\}$ should satisfy the Bethe equations. 
The pseudovacuum $|0\rangle$ ($\langle 0| \equiv |0\rangle^\dagger$) has the following properties:
\begin{eqnarray}
&\langle 0 | A(\lambda) = a(\lambda) \langle 0 | \qquad &A(\lambda) |0\rangle = a(\lambda) |0\rangle\\
&\langle 0 | D(\lambda) = d(\lambda) \langle 0 | \qquad &D(\lambda) |0\rangle = d(\lambda) |0\rangle\\
&\langle 0 | B(\lambda) = 0 \qquad &C(\lambda) |0\rangle = 0.
\end{eqnarray}
The form of the vacuum eigenvalues $a(\lambda)$ and $d(\lambda)$ depend on the specific representation of the $R$-matrix.

In order to describe the geometric quench we decompose the Hilbert space  into two spatial intervals:
one of length $L_1$ and one of length $L_2-L_1$ (the part that is added):
%$\{1 \ldots L_1\}$ and $ \{L_1+1 \ldots L_2\} $: 
$\mathcal{H} = \mathcal{H}_1 \otimes \mathcal{H}_2$.  The operators before the quench act on $\mathcal{H}_1$, while the operators after the quench act on $\mathcal{H}$.
To be able to make a connection between the two we consider the generalized two-site model introduced in \cite{Izergin1984}, 
namely the monodromy matrix $T(\lambda)$ of any Algebraic Bethe Ansatz solvable model can be written as a matrix product of two factors
\begin{equation}\label{twosite}
T(\lambda) = T_2(\lambda) T_1(\lambda)
\end{equation}
with $T_i(\lambda) \in V_0 \otimes \mathcal{H}_i$. 
From the decomposition of the monodromoy matrix follows that
\begin{equation}
B(\lambda) = A_2(\lambda) B_1(\lambda) + B_2(\lambda) D_1(\lambda)
\end{equation}
and similar expressions for the $A(\lambda)$, $C(\lambda)$ and $D(\lambda)$ operators.
We can write the initial $N$-particle state (at $t=0^+$) as
\begin{equation}
\langle \Psi^{(1)}(\{\lambda\})| =   \left(\;_1\langle 0|  \prod_{j=1}^N C_1(\lambda_j)\right)  \otimes \;_2 \langle 0|,
\end{equation}
while eigenstates in $\mathcal{H}$ can be expressed as
\begin{equation}
\hspace{-2cm}
|\Psi^{(2)}(\{\mu\}) \rangle = \left( \prod_{j=1}^N B_1(\mu_j) |0\rangle_1  \right) \otimes \left( \prod_{j=1}^N A_2(\mu_j) |0\rangle_2  \right) + \mbox{terms without overlap.}
\end{equation}
The terms without overlap are terms where not all $B_1(\mu)$ act on $|0\rangle_1$ and will therefore not contribute to the decomposition of the initial state.
The overlaps then become
\begin{eqnarray}\label{QOverlapPBC}\nonumber
\langle \Psi^{(1)}(\{\lambda\})|\Psi^{(2)}(\{\mu\})\rangle 
&= \prod_{j=1}^N d_1(\mu_j) \;_1\langle 0| \prod_{k=1}^N  C_1(\lambda_k) \prod_{l=1}^N B_1(\mu_l) |0\rangle_1\\
&= \prod_{j=1}^N d_1(\mu_j)\langle \Psi^{(1)}(\{\lambda\})|\Psi^{(1)}(\{\mu\})\rangle.
\end{eqnarray}
The scalar product $\langle \Psi^{(1)}(\{\lambda\})|\Psi^{(1)}(\{\mu\})\rangle$ is of a special type namely, the state  $\langle \Psi^{(1)}(\{\lambda\})|$ is an eigenstate while  $|\Psi^{(1)}(\{\mu\})\rangle$ is not. This type of scalar product can be represented efficiently as a determinant \cite{Slavnov1989,Maillet1999}.  To obtain the normalized quench overlap  $F_1$ as given in \eref{overlapF1} one should divide \eref{QOverlapPBC} by the norms of the states \cite{Gaudin1972, Gaudin1981,Korepin1982} before and after the quench. The quench overlap is now defined up to a momentum dependent phase, which is related to the representation of the $R$-matrix.

So far we considered periodic boundary conditions, however we can easily generalize this by considering twisted boundary conditions. In this case  the scalar product is given in \cite{Maillet2005}. Since the twist of the state after the quench does not appear explicitly, one has the freedom to chose the twists before and after the quench independently.

\subsection{Geometric quench overlap: Open boundary conditions}
The geometric quench setup has a natural generalization to the perhaps more physical case of open boundary conditions.

For an integrable system with open boundary conditions one should define the boundary conditions in terms of matrices $K_-(\lambda)$ and $K_+(\lambda)$. 
For XXZ-like models the diagonal solution of the boundary Yang-Baxter equation \cite{Cherednik1984} is given by
\begin{equation}
K_{\pm}(u) = \left( \begin{array}{cc} 
\varphi(u\pm \eta \pm \xi_{\pm}) & 0\\
0&\varphi(-u\pm \eta \pm \xi_{\pm})\,
\end{array} \right)
\end{equation}
with $K_-$ and $K_+$ describing the left and right boundaries respectively.  Each is parametrized by $\xi_-,
\xi_+$. Following \cite{Sklyanin1988} we  introduce `double-row' monodromy matrices $\mathcal{U}_-$ and $\mathcal{U}_+$:
\begin{equation}
\mathcal{U}_-(\lambda) = T(\lambda) K_-(\lambda)\widehat{T}(\lambda) = \left(\begin{array}{cc} \mathcal{A}_-(\lambda) & \mathcal{B}_-(\lambda)\\
\mathcal{C}_-(\lambda) &\mathcal{D}_-(\lambda)
\end{array}\right)
\end{equation}
\begin{equation}
\mathcal{U}_+^{t_0}(\lambda) = T(\lambda)^{t_0} K_+^{t_0}(\lambda)\widehat{T}(\lambda)^{t_0} =  \left(\begin{array}{cc}\mathcal{A}_+(\lambda) & \mathcal{B}_+(\lambda)\\
\mathcal{C}_+(\lambda) &\mathcal{D}_+(\lambda)
\end{array}\right)
\end{equation}
where
\begin{equation}
\widehat{T}(\lambda) = (-1)^L  \sigma_0^y T^{t_0}(-\lambda) \sigma_0^y
\end{equation}
and $t_0$ denotes matrix transposition in the auxiliary space.
The operators $\mathcal{A}_\pm(\lambda),\mathcal{B}_\pm(\lambda),\mathcal{C}_\pm(\lambda)$ and $\mathcal{D}_\pm(\lambda)$  are called the boundary operators in contrast with $A(\lambda),B(\lambda),C(\lambda)$ and $D(\lambda)$ which are called the bulk operators. Eigenstates can be constructed using boundary operators
\begin{equation}
\langle \Psi_\pm(\{\lambda\})| = \langle 0 | \prod_{k=1}^N \mathcal{C}_{\pm}(\lambda_j), \qquad  |\Psi_{\pm}(\{\lambda\}) \rangle = \prod_{k=1}^N \mathcal{B}_{\pm}(\lambda_j) |0\rangle.
\end{equation}
The action of boundary operators on the vacuum is similar to that of the bulk operators, expect for different vacuum eigenvalues.

We now consider a geometric quench where we keep the left boundary fixed and move the right boundary from $L_1$ to $L_2$. This allows us to write  everything in terms of  $\mathcal{B}_-(\lambda)$ and $\mathcal{C}_-(\lambda)$ operators.  Equivalently we could fix the right boundary and write everything in terms of $\mathcal{B}_+(\lambda)$ and $\mathcal{C}_+(\lambda)$. Similar to the case of periodic boundary conditions we factorize the monodromy matrix as
\begin{eqnarray}\nonumber
\mathcal{U}_-(\lambda) &= T_2(\lambda) T_1(\lambda) K_-(\lambda) \widehat{T}_1(\lambda) \widehat{T}_2(\lambda)\\
&= T_2(\lambda) \mathcal{U}_{-,1} \widehat{T}_2(\lambda)
\end{eqnarray}
where $\mathcal{U}_{-,1}$ is the double-row monodromy of the smaller system with the same left boundary. We can now decompose the $\mathcal{B}_{-}(\lambda)$ operator as a combination of bulk and boundary operators:
\begin{eqnarray}\nonumber
\mathcal{B}_-(\lambda) &= \left(A_2(\lambda) \mathcal{B}_{-,1}(\lambda) + B_2(\lambda)  \mathcal{
D}_{-,1}(\lambda)  \right) A_2(-\lambda)\\ &- \left(  A_2(\lambda)  \mathcal{A}_{-,1}(\lambda)    + B_2(\lambda)  \mathcal{C}_{-,1}(\lambda)     \right)B_2(\lambda).
\end{eqnarray}
We write the initial state as
\begin{equation}\label{BoundaryInitialState}
\left( \;_1 \langle 0 | \prod_{k=1}^N  \mathcal{C}_{-,1}(\lambda_k)  \right) \otimes \:_2 \langle 0|,
\end{equation}
and we write the final state as
\begin{equation}
\hspace{-2cm}  \left(  \prod_{k=1}^N\mathcal{B}_{-,1}(\mu_k) |0\rangle_1 \right)  \otimes  \left( \prod_{k=1}^N A_2(\mu_k)A_2(-\mu_k) |0\rangle_2 \right) + \mbox{terms without overlap}
\end{equation}
The terms without overlap are terms with no $\mathcal{B}_{-,1}$ operator acting on $|0\rangle_1$ and will not contribute when taking the scalar product with \eref{BoundaryInitialState}. This results in the geometric quench overlap:
\begin{eqnarray}\nonumber
\hspace{-2cm}\langle \Psi^{(1)}_{-}(\{\lambda\})|\Psi^{(2)}_{-}(\{\mu\})\rangle  &=  \prod_{j=1}^N a_2(\mu_j) a_2(-\mu_j) \;_1\langle 0| \prod_{k=1}^N \mathcal{C}_-(\lambda_k) \prod_{l=1}^N \mathcal{B}_-(\mu_l) |0\rangle_1\\ 
&=  \prod_{j=1}^M a_2(\mu_j) a_2(-\mu_j)\langle \Psi^{(1)}_{-}(\{\lambda\})|\Psi^{(1)}_{-}(\{\mu\})\rangle.
\label{QmatrixOPC}
\end{eqnarray}
For the XXZ spin chain, the determinant representation for 
the scalar product $\langle \Psi^{(1)}_{-}(\{\lambda\})|\Psi^{(1)}_{-}(\{\mu\})\rangle$ is computed in \cite{Wang2002,Maillet2007a}  and is related to the overlap $F_1$ of \eref{overlapF1} 
by a normalization prefactor \cite{Wang2002,Maillet2007a}. Since \eref{QmatrixOPC} does not depend on the right boundary condition, we can not only expand the system but also change the right boundary condition after the quench.

\section{Conclusions and perspectives}
We have shown how a sudden, global change in the geometrical support of an integrable model could be completely
described within Bethe Ansatz, by providing an exact decomposition of the original (arbitrarily chosen) eigenstate
into the basis of eigenstates after such a geometric quench.  The idea is quite simple, but potentially very powerful:
it applies directly to any model for which the Slavnov formula for wavefunction overlaps is explicitly known,
and allows at least in principle to reconstruct the exact subsequent time evolution of 
wavefunctions, operator expectation values and time-dependent correlation functions via techniques similar to the
ones used in the equilibrium case.  On the practical side, since the energy of the states populated by the geometric quench
can be obtained nonperturbatively, results can be accurately given for arbitrarily long times after the quench.
Moreover, since the space of eigenstates is mapped out in detail via the Bethe Ansatz, extremely efficient truncations
of the required Hilbert space sums can be implemented.  We will explore a number of specific cases in forthcoming publications.

\section*{Acknowledgements}  
We thank N. Slavnov for useful discussions.  
All authors gratefully acknowledge support from the FOM foundation of the Netherlands,
and from the INSTANS activity of the ESF.
\\

\end{document}